\documentclass[authoryear,preprint,review, 12pt]{elsarticle}

\usepackage{amssymb}
\usepackage{amsthm}
\usepackage{amsmath}
\usepackage{amsfonts}
\usepackage{amsbsy}
\usepackage{natbib}
\usepackage{bm}

\usepackage[utf8]{inputenc}
\usepackage[american]{babel}

\usepackage{color}
\definecolor{gray}{rgb}{0.5,0.5,0.5}

\journal{Statistics and Probability Letters}

\def\Mo{{\mathcal M}_0}
\def\M1{{\mathcal M}_1}
\def\Mi{\mathcal M_i}
\def\hth{\hat\theta}

\begin{document}

\begin{frontmatter}

%% Title, authors and addresses

%% use the tnoteref command within \title for footnotes;
%% use the tnotetext command for theassociated footnote;
%% use the fnref command within \author or \address for footnotes;
%% use the fntext command for theassociated footnote;
%% use the corref command within \author for corresponding author footnotes;
%% use the cortext command for theassociated footnote;
%% use the ead command for the email address,
%% and the form \ead[url] for the home page:
%% \title{Title\tnoteref{label1}}
%% \tnotetext[label1]{}
%% \author{Name\corref{cor1}\fnref{label2}}
%% \ead{email address}
%% \ead[url]{home page}
%% \fntext[label2]{}
%% \cortext[cor1]{}
%% \address{Address\fnref{label3}}
%% \fntext[label3]{}

\title{Testing Order Constraints: Qualitative Differences Between Bayes Factors and Normalized Maximum Likelihood}

%% use optional labels to link authors explicitly to addresses:
%% \author[label1,label2]{}
%% \address[label1]{}
%% \address[label2]{}

\author{Daniel W. Heck\corref{cor1}}
\ead{dheck@mail.uni-mannheim.de}
\address{University of Mannheim}
\cortext[cor1]{The authors thank two anonymous reviewers for their helpful comments and suggestions. The project was supported partly by grant 283876 from the European Research Council. Correspondence concerning this article may be addressed to Daniel W. Heck, University of Mannheim, Department of Psychology, Schloss EO 254, 68131 Mannheim, Germany, to Eric-Jan Wagenmakers, University of Amsterdam, Department of Psychological Methods, Weesperplein 4, 1018 XA Amsterdam, the Netherlands, or to Richard Morey, Cardiff University, School of Psychology, 70 Park Place, Cardiff CF10, UK.}

\author{Eric-Jan Wagenmakers\corref{cor1}}
\ead{ej.wagenmakers@gmail.com}
\address{University of Groningen}

\author{Richard D. Morey\corref{cor1}}
\ead{richarddmorey@gmail.com}
\address{Cardiff University}

\begin{abstract}
%\40 WORDS: #now 37
We compared Bayes factors to normalized maximum likelihood for the simple case of selecting between an order-constrained versus a full binomial model. This comparison revealed two qualitative differences in testing order constraints regarding data dependence and model preference.
\end{abstract}


\end{frontmatter}

\section{Model Selection with Bayes Factors and Normalized Maximum Likelihood}
\label{intro}
% model selection intro

Although all model selection methods address the inevitable trade-off between goodness-of-fit and complexity, the manner in which they measure and penalize model complexity can differ substantially. In popular information criteria such as AIC or BIC, model complexity is measured solely by the number of free parameters. Alternative approaches reflect a more subtle view on model complexity and consider --explicitly or implicitly-- not only the dimensionality of a model, but also order constraints on parameters and their functional form. Here we compare two model comparison methods that are based on very different statistical philosophies: Bayes factors for belief revision and normalized maximum likelihood for data compression.

% Bayes factors
The first method under consideration, the Bayes factor, is defined as the ratio of two marginal likelihoods \citep{Kass1995}:
\begin{equation}
\text{B}_{01}=\frac {p(y\mid\Mo)}  {p(y\mid\M1)},
\label{b01}
\end{equation}
where the marginalization occurs over the prior distribution, $p(y\mid\Mi)=\int_\Theta p(y \mid \theta , \Mi)p(\theta \mid \Mi)\,\text{d}\theta$. Complex models make many predictions; by averaging the adequacy of these predictions for the observed data over the prior, the Bayes factor automatically and implicitly penalizes for model complexity. An order constraint results in a larger marginal probability if it reduces the parameter space to areas of high likelihood. From the perspective of belief revision, Bayes factors measure the extent to which the data mandate a change from prior to posterior model odds. As such, Bayes factors represent ``the standard Bayesian solution to the hypothesis testing and model selection problems" \citep[p. 648]{LewisRaftery1997}.

% minimum description length and BF
The second method under consideration, normalized maximum likelihood, is an instantiation of the minimum description length (MDL) principle \citep{Rissanen1978, Grunwald2007}. According to MDL, a statistical model may be interpreted as a method to compress data. If a model captures structural patterns in a data set, it can be used for compressing that data set, resulting in a shorter code length. However, the model itself also has to be encoded, thereby inducing a premium on parsimony. The solution to the problem of finding the optimal encoding is to select the model with the largest normalized maximum likelihood \citep[NML;][]{Rissanen2001},
\begin{equation}
\text{NML}_i = \frac {p(y \mid \hth_{y,i})}{ \int_\mathcal X p(x \mid \hth_{x,i} ) \,\text{d}x},
\label{nml}
\end{equation}
where $\hth_{y,i}$ is the maximum likelihood (ML) estimator for data $y$ and model $\Mi$. The normalizing integral in (\ref{nml}) ranges over the entire sample space $\mathcal{X}$; hence, NML measures complexity explicitly, by integrating over the sample space, and models are punished to the extent that they are able to provide a good fit to a wide range of possible observations. Adding order constraints to a model reduces the fit in some areas of the data space and thereby results in a smaller penalty term.

% Luckiness
Often, the normalizing integral in (\ref{nml}) is not defined. As a solution, the more general luckiness NML \citep[LNML; ][p. 309]{Grunwald2007} was developed, in which the likelihood function $p(y \mid \theta)$ in (\ref{nml}) is replaced by the weighted likelihood 
\begin{equation}
p^L(y \mid \theta) = p(y \mid \theta) e^{-a(\theta)},
\end{equation}where $a(\theta)$ is a continuous luckiness function.
%\footnote{Specifically, we use LNML-2 where the estimator of $\theta$ is defined as $\hth=\arg\max_\theta [p(y \mid \theta)e^{-a(\theta)}]$ (in contrast to LNML-1 with the ML estimator  $\hth=\arg\max_\theta p(y \mid \theta)$.} 
This function specifies subspaces of the parameter space where model selection by means of LNML will be more efficient (i.e., one might `get lucky' in compressing the data). Note that LNML reduces to the standard NML if a constant, nonzero luckiness function $a(\theta)=c$ is used.

% asymptotic results and differences
Model selection by NML is asymptotically indistinguishable from model selection by Bayes factors with Jeffreys' prior \citep{Rissanen1996}. Moreover, with the introduction of LNML, it is possible to define luckiness functions for LNML that match the priors of Bayes factors and will yield identical asymptotic results \citep[][p. 313]{Grunwald2007}. For some statistical models such as one-dimensional Gamma or Gaussian models, multiple regression, and Gaussian process models, the two methods yield identical results for all sample sizes \citep{Bartlett2013, Kakade2006}.
 
However, the philosophy that underlies the two approaches is markedly different. Whereas MDL aims at data compression, the Bayes factor is concerned with belief revision. Moreover, in LNML, the complexity of a model is defined as an explicit value (as an integral over the sample space), independent of the data set under consideration. In contrast, Bayes factors consider complexity implicitly by integrating the adequacy of a model's predictions for the observed data across the parameter space, weighted by the prior.

Here, we show how these general differences between Bayes factors and NML are reflected in two specific qualitative differences when testing order constraints. Insights about the way how both methods account for order constraints are important because many information criteria cannot be used for this kind of problem. Specifically, we provide an existence proof by considering a simple test for an order constraint on a binomial rate parameter.

\section{Example: Evaluating an Order Constraint for a Binomial Rate Parameter}
\label{models}
% models
Under the full model $\mathcal M_1$, $N$ binary observations are assumed to be binomially distributed with rate parameter $\theta$, that is, $y \sim \text{Bin}(N,\theta)$. The competing model $\mathcal M_0$ has the additional order constraint $\theta \leq z$ for a fixed value $z \in (0,1)$. Note that both models feature a single free parameter, necessitating the use of a model comparison approach that measures complexity by more than just the number of free parameters.

% derive Bayes factors
\subsection{Bayes Factor}
\label{Sbf}
We assign $\theta$ a uniform prior under both models $\mathcal M_0$ and $\mathcal M_1$. Because the priors for $\theta$ under both models are proportional for $\theta \leq z$, the Bayes factor in favor of the constraint can be computed as the ratio of posterior to prior mass of the full model $\mathcal M_1$ over the range $\theta \in [0,z]$ \citep{Klugkist2007}:
\begin{eqnarray}
\text{B}_{01} &=&  \frac{\int_0^z p(\theta \mid  y, \M1) \,\text{d}\theta }{\int_0^z p(\theta \mid \M1) \,\text{d}\theta}
\label{bf.calc}\\
        &=&  \frac 1 {z \text{Be}(y+1,N-y+1)} \int_0^z\theta^y (1-\theta)^{N-y} \,\text{d}\theta, \label{bf}
\label{bf4}
\end{eqnarray}
where Be(\textit{a,b}) denotes the beta function. With equal prior odds, the posterior model probability in favor of the constrained model is
\begin{equation}
w_0^\text{B} = \frac {\text{B}_{01}} {1+ \text{B}_{01}}.
\label{wbf}
\end{equation}

% derive NML
\subsection{Luckiness Normalized Maximum Likelihood}
\label{Snml}

According to \citet[][p. 313]{Grunwald2007}, LNML with the luckiness function $a(\theta)$ is asymptotically identical to the Bayes factor with prior $p(\theta \mid \mathcal M)$ if
\begin{equation}
p(\theta \mid \mathcal M ) \propto \sqrt{\det \mathcal I(\theta)}e^{-a(\theta)},
\end{equation}
where $\mathcal I(\theta)$ is the Fisher information. For the two binomial models under scrutiny, $\mathcal I(\theta) = \theta^{-1} (1-\theta)^{-1}$ and hence the luckiness function
\begin{equation}
a(\theta ) = -\ln \theta^{1/2} (1-\theta)^{1/2}
\label{luckiness}
\end{equation}
matches the uniform prior in (\ref{bf4}) for both models.

For our simple scenario, the discrete sample space $\mathcal{X}$ can easily be enumerated to compute the LNML normalizing integral 
\begin{equation}
 \int_\mathcal X p(x \mid \hth^L_{x,i} )\exp({-a(\hth^L_{x,i})}) \,\text{d}x.
\end{equation}
Specifically, the LNML normalizing integral equals the sum of the weighted likelihood values for all possible data sets in $\mathcal{X}$. The estimator $\hth^L_{y,1}$ of the full model maximizes the luckiness-weighted likelihood $p^L(y \mid \theta)$, 
\begin{align}
\hth^L_{y,1}&=\arg\max_\theta [  \theta^{y+ 1/2}(1-\theta)^{n-y+1/2}]\nonumber\\
&=\frac{y+1/2}{n+1},
\end{align} 
and is identical to that of the constrained model if  $\hth^L _{y,1} \leq z$. Otherwise, the order constraint is violated and $\hth^L _{y,0}=z$.

As a measure of the degree to which LNML prefers a model over its competitors, the probability of model $i$ being the best model at hand can be computed using LNML model weights,
\begin{equation}
w_i^\text{L}=\frac{ \text{LNML}_i}{\sum_j \text{LNML}_j}.
\label{wnml}
\end{equation}
The model weights $w_i^\text{L}$ are conditional on the data and are analogous to posterior model probabilities. Therefore, they provide a way to directly compare model preference between Bayes factors and LNML. Note that the two qualitative differences emerge for both LNML and standard NML.\footnote{The supplements show the comparison between standard NML and the asymptotically identical Bayes factor based on Jeffreys' prior.} 

\section{Results: Qualitative Differences Between Bayes Factors and LNML}

\subsection{Data Dependence}
\label{datadependent}

\begin{figure}[ht!]
	\centering
	\includegraphics[width=.8\textwidth]{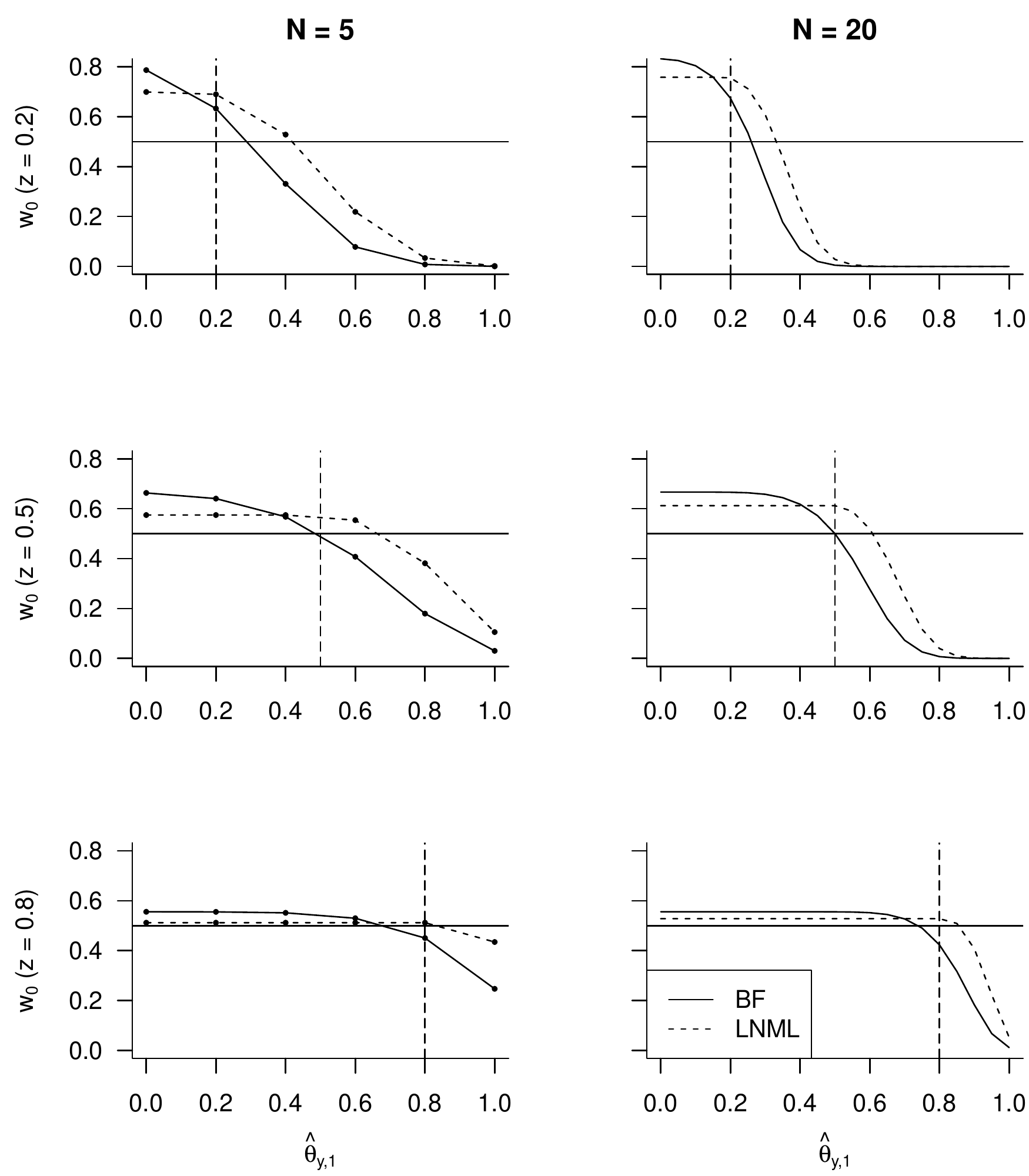}
	\caption{In contrast to the Bayes factor, LNML model selection is independent of the observed data if the order constraint is satisfied, as shown by constant model weights $w_0$ in the range of $\hth_{y,1} \leq z$. The boundary $z$ is shown as a vertical, dashed line. For $N=20$, dots for discrete observations are omitted. Weights that exceed the horizontal line ($w_0=0.5$) indicate a preference for the constrained model.}
	\label{Ftheta}
\end{figure}

If the estimator of the full model $\mathcal{M}_1$ satisfies the order constraint of $\mathcal{M}_0$ (i.e., $\hth^L_{y,1} \leq z$), the numerator $p^L(y \mid \hth^L_{y,i})$ of LNML in (\ref{nml}) is identical for both models. In this situation, LNML model selection no longer depends on the observed data $y$, since
\begin{equation}
w_0^\text{L} = \frac{\int_\mathcal{X} p^L(x \mid \hth^L_{x,1}, \M1 )\,\text{d}x} {\int_\mathcal{X} p^L(x \mid \hth^L_{x,0}, \Mo )\,\text{d}x + \int_\mathcal{X} p^L(x \mid \hth^L_{x,1}, \M1 )\,\text{d}x}.
\label{independence}
\end{equation}
Note that this result holds for testing order constraints in general and is not restricted to the binomial model. Figure \ref{Ftheta} shows how the model weight $w_0^\text{L}$ changes depending on the observed data. The data independence of LNML results in a constant model weight whenever $\hth^L_{y,1} \leq z$. In contrast, model selection by the Bayes factor is always sensitive to the observed data, including data with $\hth^L_{y,1} \leq z$. In such cases, the more the constraint is satisfied, the larger the Bayes factor in favor of the restriction becomes.

\begin{figure}[ht!]
	\centering
	\includegraphics[width=\textwidth]{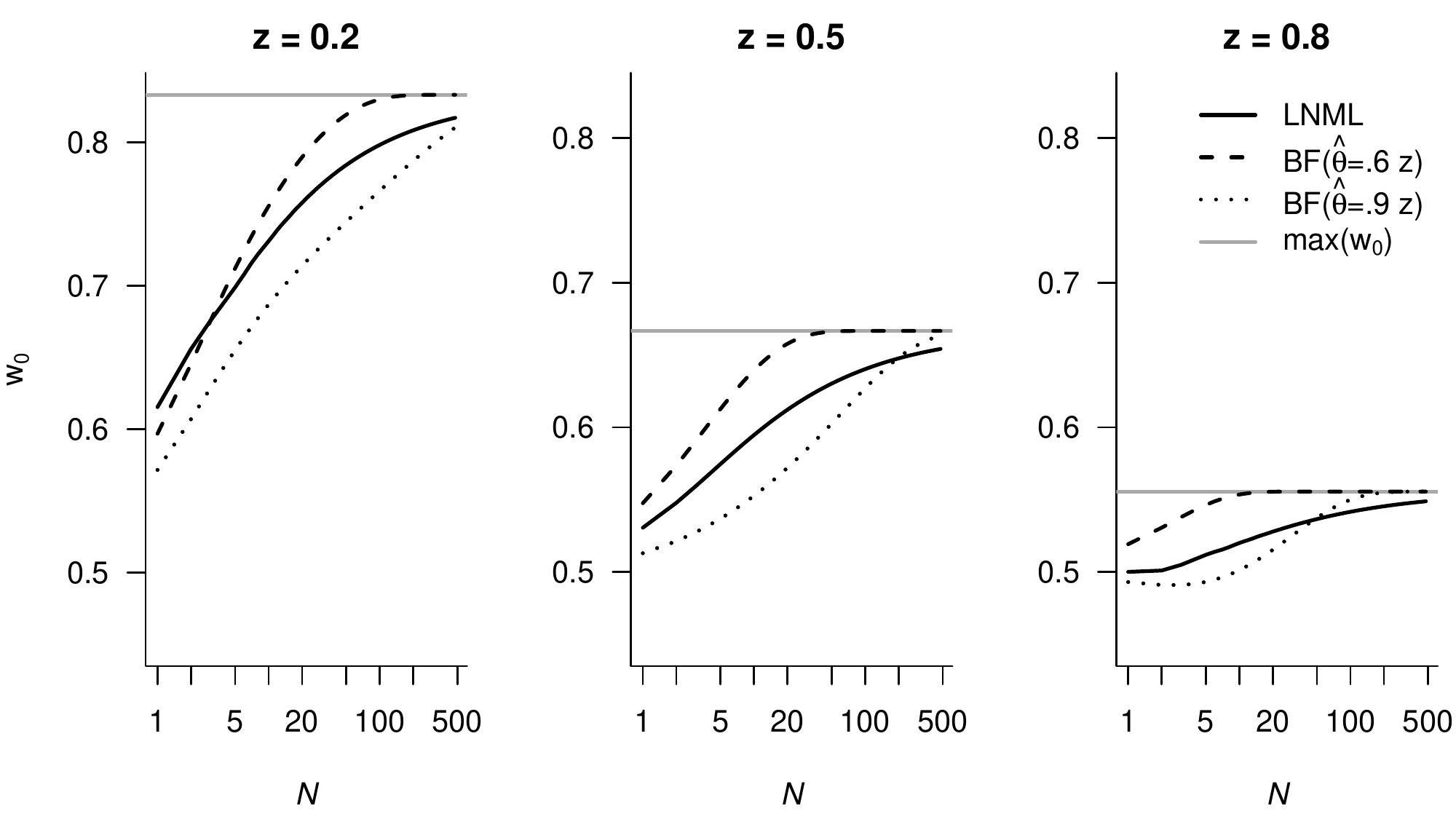}
	\caption{In contrast to LNML, the convergence rate of the Bayes factor to the maximum model weight in favor of the order constraint depends on the exact data if $\hth_{y,1} \leq z$. The two Bayes factors shown correspond to data resulting in ML estimates of $\hth_{y,1} = 0.6 z$ and $\hth_{y,1} = 0.9 z$ for all $N$.}
	\label{Fn}
\end{figure}

The data dependence of the Bayes factor results in different convergence rates to the maximum possible weight in favor of the order constraint. For the Bayes factor, it follows from (\ref{bf.calc}) that under uniform priors on $\theta$,
\begin{equation}
z \text{B}_{01} = \int_0^z p(\theta \mid y, \M1) \,\text{d}\theta < 1,
\end{equation}
and thus, that $B_{01} < 1/z$. Accordingly, the maximum posterior probability in favor of the order constraint is $w_0^\text{B} = 1/(1+z)$. Given that the constraint holds, the Bayes factor will converge to this model weight. However, Figure \ref{Fn} shows that the speed of convergence to this maximum depends on the exact data. For instance, if $\hth_{y,1}=.9z$, larger samples are required to obtain evidence in favor of the order constraint compared to less ambiguous data with $\hth_{y,1}=.6z$. Because of the matching luckiness function $a(\theta)$ in (\ref{luckiness}), LNML converges to the same maximum model weight. However, if the order constraint is satisfied, the speed of convergence does not depend on the exact data. For unambiguous data, LNML might therefore require larger samples to support the order constraint than the Bayes factor.

In sum, whenever the estimator for the constrained model equals that of the full model (i.e., the order constraint is satisfied), LNML no longer depends on the observed data. In contrast, the Bayes factor remains sensitive to the observed data.

\subsection{Model Preference}
\begin{figure}[ht!]
% \centering
\includegraphics[width=\textwidth]{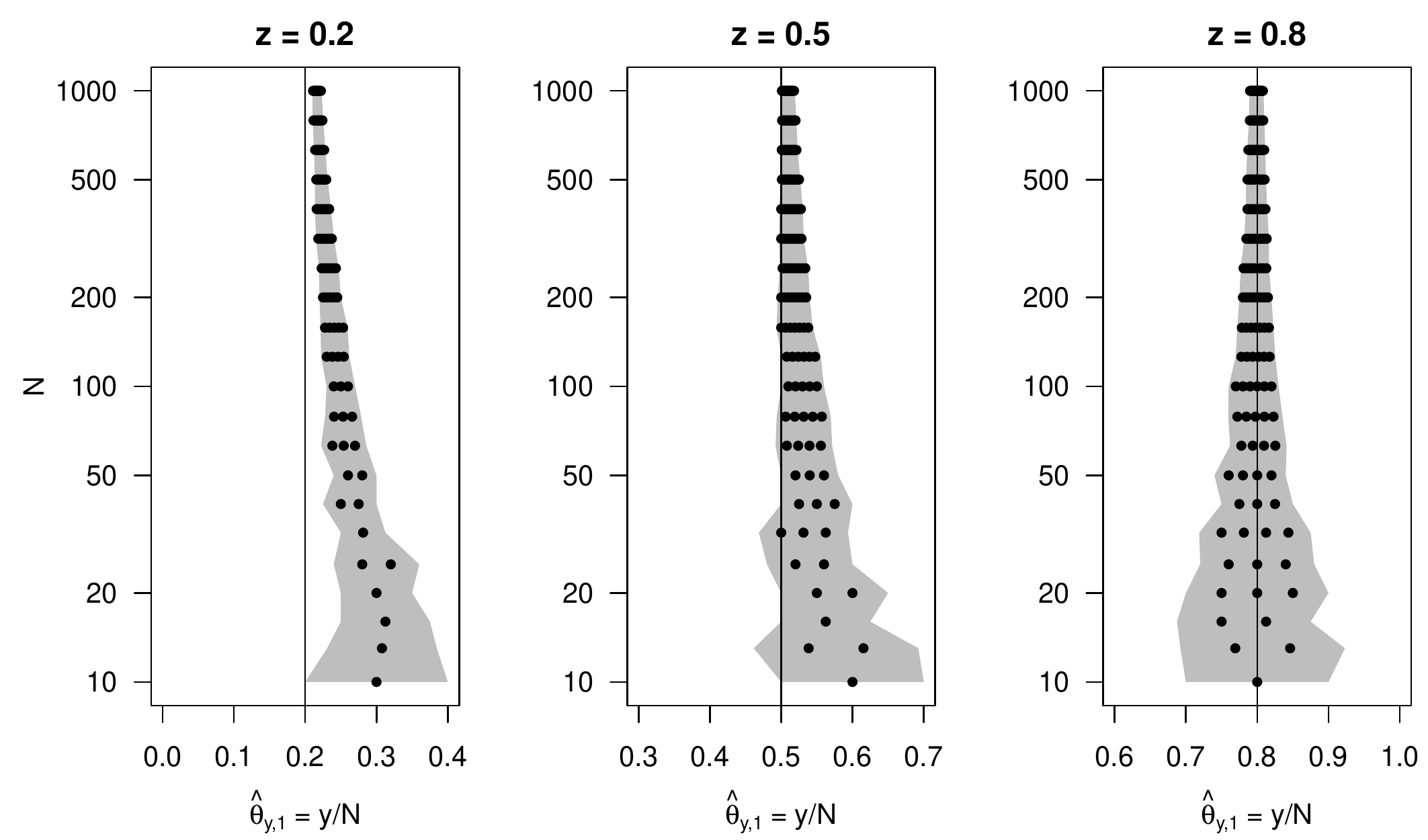}
\caption{Black dots represent possible data for which LNML prefers the order-constrained model, whereas the Bayes factor favors the full binomial model. The gray area shows divergence in model preference and is bounded to the left and the right by data for which the Bayes factor and LNML agree which model to prefer.}
\label{FthetaN}
\end{figure}

\begin{figure}[ht!]
% \centering
\includegraphics[width=8cm]{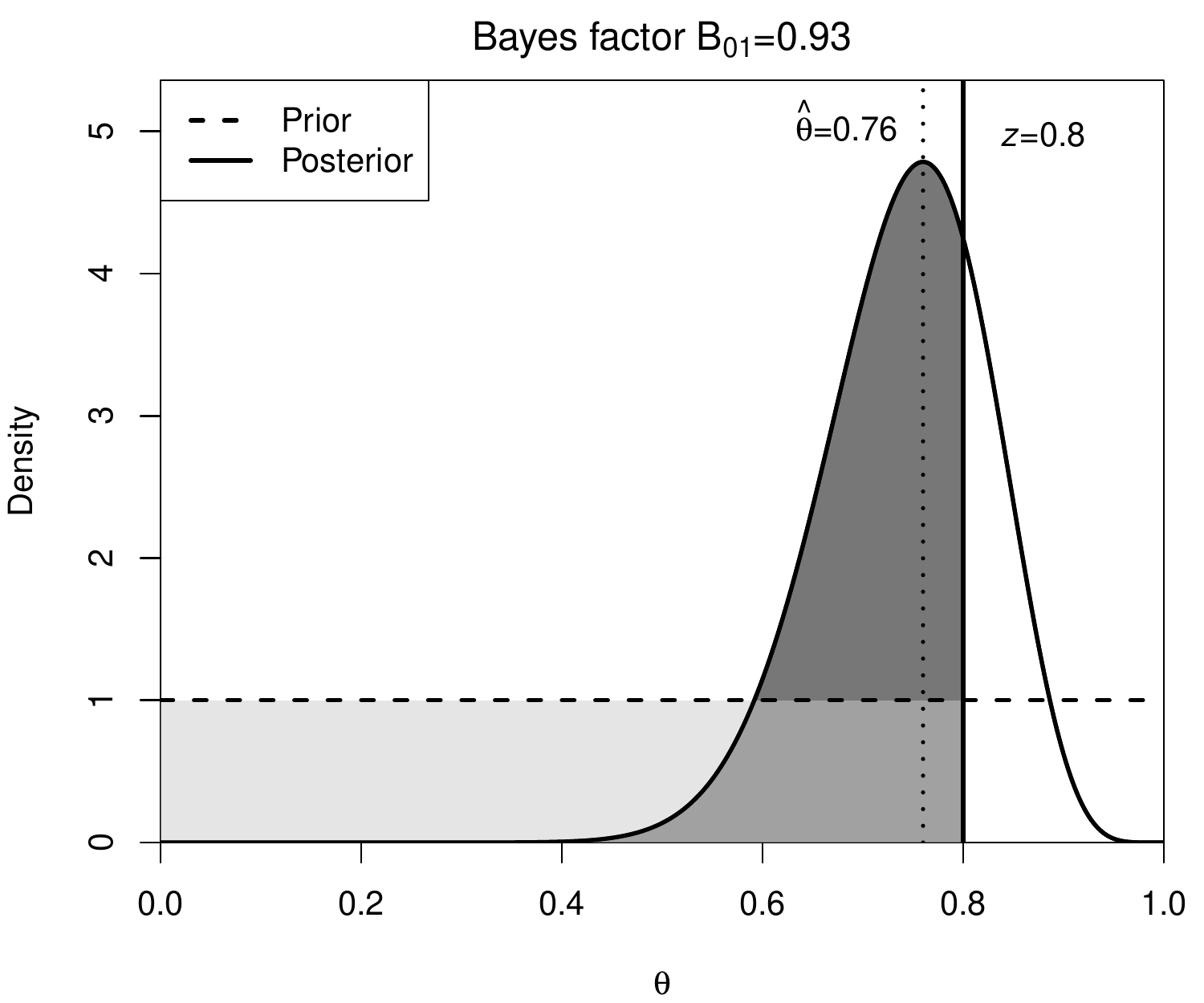}
\caption{The Bayes factor can prefer the unconstrained model even though the order constraint $\theta \leq z$ is satisfied by the ML estimator ($N=25$, $y=19$). $\text{B}_{01}$ is computed as the ratio of posterior mass (dark gray) to prior mass (light gray) on the order-restricted range $[0,z]$.}
\label{FratioBF}
\end{figure}

Figure \ref{FthetaN} shows data for which the Bayes factor and LNML prefer a different model. In these cases, LNML selects the constrained model, whereas the Bayes factor based on uniform priors prefers the full model. For a boundary of $z = 0.8$, for instance, the Bayes factor sometimes prefers the full model even though the ML estimator satisfies the order constraint. This occurs when the posterior for $\theta$ under $\mathcal{M}_1$ has less mass over the range $\theta \leq .8$ than the prior for $\theta$ under $\mathcal{M}_1$ (cf. Eq. \ref{bf4}). Figure \ref{FratioBF} illustrates this counterintuitive result.

The proportion of possible data sets with diverging results increases with $z$. For example, with $N=20$ and $z=0.2$, only $5$\% of possible data sets lead to diverging model preferences, increasing to $10$\% for $z=0.5$ and $15$\% for $z=0.8$. However, for larger sample sizes, the differences in model preference between Bayes factors and LNML decrease; for example, when $N=1000$, the proportions of critical data sets fall to $1.2$\%, $1.8$\%, and $1.9$\%, respectively. Note that in all of these critical cases, the model weights of both methods only show weak preferences for or against the order constraint (i.e., all $w_0^L<0.77$ and all $w_0^B>0.16$). Therefore, this qualitative difference might only have minor consequences for model selection in practice.

In sum, for most data sets both LNML and the Bayes factor will prefer the same model even in small samples. However, for ambiguous data where the ML estimator is near the order constraint, LNML and the Bayes factor may prefer different models.

\section{Discussion}
% summary
We identified two qualitative differences between Bayes factors and LNML in testing order constraints, which also apply to standard NML as a special case. First, if the order constraint is satisfied, LNML is independent of the observed data, whereas the Bayes factor remains dependent on the observed data. This implies that the speed of evidence accumulation in favor of an order constraint is constant for LNML, but depends on how well the constraint is satisfied for the Bayes factor. Second, in some cases, the Bayes factor may favor the full model while LNML prefers the constrained model. Whereas the data independence of LNML holds for tests of order constraints in general (cf. Eq. \ref{independence}), differences in model preference might depend on the exact model. However, we expect that preferences are more likely to differ close to the boundary and decrease for larger samples, similarly as for the binomial model.

% conclusion
One common advantage of Bayes factors and NML concerns their ability to take order constraints into consideration, contrary to model selection tools such as AIC or BIC. Although several authors have stressed the similarities between Bayes factors and NML 
\citep[e.g.,][]{Grunwald2007, Bartlett2013, Kakade2006}, a detailed study of order-constrained inference shows that what is good for belief revision (Bayes) is not necessarily good for data compression (NML).

%% If you have bibdatabase file and want bibtex to generate the
%% bibitems, please use
%%
% \section{Literature}
\bibliographystyle{elsarticle-harv}
\bibliography{referenties,lab}

%% The Appendices part is started with the command \appendix;
%% appendix sections are then done as normal sections
\appendix

\end{document}